\documentstyle[aps,prl,epsf,floats,amsbsy]{revtex}

\def\boldtau{{\boldsymbol{\tau}}}
\def\boldr{{\bf r}}
\def\boldk{{\bf k}}
\def\boldR{{\bf R}}
\def\bold0{{\bf 0}}
\def\unitr{\hat{\bf r}}
\def\Schrodinger{Schr\"odinger }

\begin{document}

\title{Tight-binding parameters from the full-potential linear muffin-tin orbital method:
A feasibility study on NiAl}
\author{David Djajaputra and Bernard R. Cooper}
\address{
Department of Physics, West Virginia University, P.O. Box 6315, Morgantown, WV 26506-6315}

\date{\today}

\maketitle

\begin{abstract}
We have examined a method of direct extraction of accurate tight-binding parameters from an
{\it ab-initio} band-structure calculation. The linear muffin-tin potential method, 
in its full-potential implementation, has been used to provide the hamiltonian and overlap 
matrix elements in the momentum space. These matrix elements are Fourier transformed to
real space to produce the tight-binding parameters. The feasibility of this method 
has been tested on the intermetallic alloy NiAl, using {\it spd} orbitals for each atom. 
The parameters generated for this alloy have been used as input to a real-space calculation 
of the local density of states using the recursion method.
\end{abstract}

\section{Introduction}

%% O(N) Methods

There has been a growing interest in recent years to construct methods of
electronic-structure calculation in which the required computational time scales 
linearly with the size of the system, the so-called $O(N)$ method.\cite{goedecker1999}
The proposed approaches abandon the \boldk-space method used in 
convential band-structure calculations in favor of working directly in real space.
\cite{yang1991,galli1992,galli2000,goedecker1994,kohn1996,yang1997,jayanthi1998}
Instead of calculating the energy bands, one instead focuses on local properties, e.g., the 
local density of states. Extensive properties, like the total energy, are obtained by 
integrating over space instead of over the Brillouin zone (hence the linear scaling with the 
size of the system). This real-space approach is particularly relevant to cases in which 
lack of perfect crystalline symmetry is an essential part of the problem (e.g., in surfaces, 
impurities, defects, and amorphous materials) since in those cases Brillouin zone and energy bands 
are simply non-existent.\cite{heine1980} The tight-binding (TB) method is among the most
popular implementations of this real-space strategy.
 
%% Tight Binding

The tight-binding approach to electronic structure has a long history which dates back to 
the classic work of Slater and Koster half a century ago.
\cite{slater1954,slater1965,sharma1979,harrison1989} 
It has been found especially useful in the modeling of the important technological material 
silicon.\cite{menon1997,bernstein1997,cohen1997,lenosky1997} The conventional means to 
obtain the TB parameters is to follow the original procedure of Slater and Koster namely to 
fit the TB energy bands to the ones obtained from accurate self-consistent calculations. 
The parameters for elemental solids obtained in this way have been compiled by 
Papaconstantopoulos.\cite{papa1986} This database has also been actively enlarged to include 
parameters for various alloys.\cite{papa1996,mehl1998,yang1998,papa1998} Recent interest 
in real-space electronic structure methods has supplied a renewed vigor to this field of 
TB parametrization.\cite{turchi1998,goringe1997,bowler1997,ohno1999,raabe1998} 

%% Critiques on Fitting

Despite its seemingly simple principle, the fitting process that one uses in obtaining
the TB parameters is not always straightforward in practice. Starting from a guess set
of TB parameters one calculates the TB energy bands which are then compared with the
accurate bands from a self-consistent calculation. One can proceed to minimize the 
merit function by using standard nonlinear optimization procedures.\cite{press1999} 
This, in general, is quite a straightforward procedure for simple systems with one atom per 
unit cell. However, for more complicated multiatom unit cells with more intricate
``spaghetti'' of bands, the number of independent parameters on which the merit function 
depends grows rapidly. In the optimization process for this case, the merit function can 
get trapped easily in a local minimum and the output TB bands have little resemblance to
the bands that they are intended to fit. Getting around this problem by choosing a good
initial set of parameters, unfortunately, still comprises more art than science. Another
drawback of the TB parameters that are obtained from fitting is that one normally fits only 
the bands below the Fermi energy and a few bands above it. In general the parameters 
obtained depend on the number of bands that are used in the fitting. This feature makes it
difficult to attach a physical meaning to the parameters.

%% Andersen's TB-LMTO

Andersen and coworkers have proposed a method to obtain TB parameters directly from his linear
muffin-tin orbital (LMTO) method.\cite{andersen1984,andersen1985,andersen1987,andersen1994,tank2000}
This method has also been extended by other groups.\cite{xie1997} The method outputs
the {\it orthogonal} hamiltonian (hopping) parameters between highly-localized (screened) 
muffin-tin orbitals in the two-center approximation. Since each hopping parameter effectively 
vanishes beyond second-nearest neighbors, TB calculations using these parameters can be highly
efficient. Moreover, the obtained parameters are also independent of
structure (transferable), a property which is highly desirable in TB applications. Most of these
nice properties, however, depend on the atomic sphere approximation (ASA) 
which is used to derive these results.\cite{andersen1984a} This approximation has been known 
to work rather well for close-packed structures but is not very accurate for open structures, 
although this can be remedied to some extent by using the so-called empty spheres.\cite{tank2000} 
The TB-LMTO method assigns a minimal base (at most one {\it s}, three {\it p}, and five 
{\it d} orbitals per atom) to each atom in the crystal. In the LMTO-ASA method, the muffin-tin 
orbitals are set to have zero kinetic energy ($\kappa^2 = 0$) in the interstitial region. 

%% Full Potential Methods

Full-potential (FP) implementations of linear electronic-structure methods, e.g., the FPLMTO method or
the FP Linear Augmented Plane Wave (FPLAPW) method, are among the most accurate 
electronic-structure methods.\cite{andersen1975,skriver1984,nemoshkalenko1998,loucks1967,singh1994}
It is therefore desirable to have a direct method of extracting practical TB parameters
from these methods. The basis functions in the FPLAPW method are solutions to 
\Schrodinger equation inside the MT sphere and plane waves in the interstitial region. Since
the basis functions are not localized, its matrix elements therefore cannot be mapped into 
the TB form. (It should be noted, however, that it is possible to project 
a FPLAPW calculation into localized basis.\cite{portal1995}) The basis functions in the 
FPLMTO method, on the other hand, can be {\it chosen} (by setting the value of
the kinetic energy $\kappa^2$) to have a decaying tail outside its MT sphere. Direct mapping
to TB parameters is therefore possible for the FPLMTO method. To our knowledge this direct approach 
has been seriously studied only by McMahan and Klepeis who used it to study Si and Si/B phases.
\cite{mcmahan1997,mcmahan1998,manh2000}

%% Motivation of this paper

In this paper we use this direct approach and extract TB (hamiltonian
and overlap) parameters for the intermetallic alloy NiAl from a FPLMTO method. These 
parameters are then used as input to a real-space calculation of the local density of 
states using the recursion method. To our knowledge, this is the first test of this direct 
method to a transition-metal alloy. Our choice of NiAl is determined by our previous 
experience with it\cite{djajaputra2001} and by the fact that NiAl has a simple B2 structure
in which each atom in the unit cell has the full cubic point group. The latter is of practical
relevance since the symmetry allows us to reduce the number of independent TB parameters 
that need to be calculated and saved in the database. Since the TB parameters
that we extract from the FPLMTO method are to be used in a recursion calculation, we do not 
need to extract the (two-center) Slater-Koster parameters as was done by McMahan and Klepeis.
\cite{mcmahan1997} Our motivation in performing this calculation is twofold. First, 
one hopes to obtain a general idea of the level of numerical accuracy that can be expected to
result from a combination method like this. Second, eventually one would like to use
the TB parameters extracted from this method in applications where it is too expensive 
computationally to use direct {\it ab-initio} methods, e.g., in studying the effects of 
a small concentration of impurity atoms to static and dynamic properties of alloys. 
Since the TB method is a {\it non}-self-consistent method, the plausibility of its 
result can only be judged by how well it reproduces certain ``benchmark'' results. In this
paper the ``benchmarks'' are the local density of states that can also be obtained from the
FPLMTO method.

%% Outline of paper

Section \ref{fplmto_section} of this paper discusses the specific implementation of the FPLMTO method
that we use for this work. The choice of basis is of the utmost importance in a TB calculation.
Since the FPLMTO basis functions are well-known for being complicated we have chosen to
discuss it in some details in Sec. \ref{basis_functions}. Practical points regarding the 
transformation to real space are listed in Sec. \ref{fourier_transform}. The recursion calculation
and the results that we obtained from it are discussed in Sec. \ref{recursion}; and major points
of the paper are discussed in a summary section at the end of the paper.

%% Section: FPLMTO

\section{FPLMTO Method}
\label{fplmto_section}
In this paper we have used an implementation of the FPLMTO method that is developed by Wills and 
Price.\cite{price1989,price1992,wills2000}. Several other implementations are available in
the literature and, although the main formalism is largely the same, each implementation 
is different in its details.\cite{nemoshkalenko1998,savrasov1992,methfessel1988,methfessel1989} 
The FPLMTO uses a fixed basis set in calculating its hamiltonian and overlap matrix elements; 
it is therefore similar to the linear combination of atomic orbitals (LCAO) method.\cite{ziman1971}
However, while in the LCAO method the basis set is determined, and fixed, at the beginning,
the basis set in the FPLMTO method is determined self-consistently. The FPLMTO method 
is very similar to the Augmented Spherical Wave (ASW) of Williams, K\"ubler, and Gelatt.
\cite{williams1979}

%% Single negative kappa

One of the strengths of the FPLMTO method is that it allows unlimited number of basis functions
to be assigned to each atom in the unit cell. This provides unlimited variational flexibility
in finding the ground state configuration. The practical requirement of TB, however, requires
us to use a minimal basis set. Furthermore, although FPLMTO method allows the use of localized
($\kappa^2 < 0$) or unlocalized ($\kappa^2 >0$) basis,\cite{springborg1987} only the localized 
basis should be used for TB purposes. Williams {\it et al.} have used $\kappa^2 \approx -0.2$ Ry
exclusively in their ASW paper.\cite{williams1979} They have also warned that the use of a 
single fixed $\kappa$ is rather restrictive and cited the work of Gunnarsson {\it et al.}
\cite{gunnarsson1976} that shows the variation of eigenenergies with $\kappa$ in a molecular 
calculation. 

\subsection{Basis Functions}
\label{basis_functions}

%% Notation

In the FPLMTO method the crystal is divided into non-overlapping muffin-tin (MT) spheres 
surrounding each atomic sites. The radius of the MT sphere that is centered at $\boldtau_\alpha$,
the position of the $\alpha$-th atom in the unit cell, is denoted by $s_\alpha$. On this 
muffin-tin geometry we construct a muffin-tin potential. This potential is spherically
symmetric within each muffin tin and equal to a constant ${\rm V_0}$, the muffin-tin zero,
at the interstitial region. The potential inside the MT, $v_{\rm MT}(r)$, does {\it not} have to
join continuously to the interstitial potential ${\rm V_0}$ at the MT radius.

%% Generalities

We emphasize that in the FPLMTO method the MT potential is used {\it solely} to construct 
the basis functions. The final output charge density and potential, which are obtained from
a self-consistent calculation, do not necessarily have 
the form of a MT potential. During each iteration in the self-consistency loop, the 
LMTO basis functions are used to solve the \Schrodinger (or Dirac) equation
in a variational manner similar to the LCAO method. 
The resulting eigenfunctions are then used to calculate the new charge density and potential.
The important difference from the LCAO method is that here, in addition to the charge density 
and potential, one also updates the basis functions on every step. This update is performed
by constructing a new MT potential from the output potential. The constant potential ${\rm V_0}$
is given the value of the average potential at the interstitial, while the spherically-symmetric 
potential $v_{\rm MT}(r)$ is obtained from the angular average of the potential 
inside the muffin tin.\cite{loucks1967,springborg2000,ham1961,mattheiss1968} 
Note that we average the potential rather than the charge density as used in 
an alternative scheme.\cite{liberman1967} 

%% More Andersen's LCMTO

The final output MT orbitals therefore provide the ``best'' set of basis functions 
(within the variational freedom imposed by the choice of the LMTO parameters) for performing an 
LCAO-like calculation on the system under consideration.
This view of the MT orbitals as basis functions for an LCAO-type calculation 
has appeared from time to time in the literature, under the name of the linear 
combination of muffin-tin orbitals (LCMTO) method. It has been particularly useful for 
calculations of the electronic structure of systems that do not possess lattice 
periodicity, e.g., surfaces, impurities, and atomic 
clusters.\cite{andersen1971,andersen1973,kasowski1976,harris1980}
Unfortunately, the difficulty in performing the necessary three-dimensional integrations in 
this method has made it less popular relative to other methods that utilize orbitals
with simpler analytical properties like the Slater- or Gaussian-type 
orbitals.\cite{springborg2000,kasowski1976,porezag1995}

%% Structure of the basis function

Each basis function in the FPLMTO method is centered around an atomic site that we will call
its host site. With respect to this site, all other atomic sites will be referred to as remote sites.
The basis function is constructed as a continuous 
patchwork of three different parts: (1) inside the host 
MT sphere; (2) at the interstitial region; (3) inside all remote MT spheres.
Thus the basis function centered at atomic site $\boldtau_\alpha$ inside the unit cell with
lattice vector $\boldR_i$ has the form:

\begin{equation}
\phi_{i \alpha}({\bf r}) = \phi_{i \alpha}(i \alpha | {\bf r}) + \phi_{i \alpha}(I | {\bf r})
+ {\sum_{i' \alpha'}}' \phi_{i \alpha}(i' \alpha' | {\bf r}).
\end{equation}

\noindent Here the primed summation means that it should be carried out over 
all $(i'\alpha') \neq (i \alpha).$ The MTO also carries other indices: $L \equiv (lm)$ which are the
angular-momentum quantum numbers of the part of the MTO inside the host sphere
$\phi_{i \alpha}(i \alpha | {\bf r})$; 
the tail parameter $\kappa$ which controls the behavior of the orbital at the interstitial 
region; and possibly also the principal quantum number $n$ which controls the number of radial
nodes of $\phi_{i \alpha}(i \alpha | {\bf r})$. The last is usually not necessary since one
normally assigns only one principal quantum number for each $(lm)$. In any case, for clarity 
we will suppress these indices unless their presence is really necessary. 

%% Radial Part

An MTO is essentially an augmented spherical wave (ASW).\cite{williams1979} This means that it
is constructed by defining it to be a spherical wave (spherical Bessel, Neumann, or 
Hankel function, i.e. solution of Helmholtz equation in spherical coordinates) in the 
interstitial region:

\begin{equation}
\phi_{i\alpha}(I | {\bf r}) = K_l(\kappa,r_{i \alpha}) \cdot i^l Y_{lm}(\hat{\bf r}_{i \alpha}) 
\cdot \Theta(I|\boldr),
\label{interstitial}
\end{equation}

\noindent with a radial part:\cite{wills2000,springborg1987}

\begin{equation}
K_l(\kappa,r) = - \kappa^{l+1} \times \cases{n_l(\kappa r) - i j_l(\kappa r), \ &$\kappa^2 < 0$, 
\cr n_l(\kappa r), \ &$\kappa^2 > 0$.}
\label{interstitial_radial}
\end{equation}

\noindent Here ${\bf r}_{i \alpha} \equiv ({\bf r} - \boldR_{i \alpha})$,
$\boldR_{i \alpha} \equiv (\boldR_i + \boldtau_\alpha)$,
$r_{i \alpha} \equiv |\boldr_{i \alpha}|$, and 
$\hat{\bf r}_{i \alpha} \equiv \boldr_{i \alpha} / r_{i \alpha}$. 
The masking function $\Theta(I|\boldr)$
is equal to 1 if $\boldr$ is in the interstitial region and 0 otherwise.
For $\kappa^2 < 0$, $\kappa = i |\kappa|$, 
this spherical wave is usually expressed in terms of the spherical Hankel 
function of the first kind:\cite{springborg1987}

\begin{equation}
K_l(\kappa,r) = i \kappa^{l+1} h_l^{(+)}(\kappa r)
= {e^{-z} \over r} \times \cases{1, &$l=0$, \cr |\kappa| (1 + z^{-1}), &$l=1$, \cr
|\kappa|^2 (1 + 3 z^{-1} + 3 z^{-2}), &$l=2$,}
\end{equation}

\noindent where $z \equiv |\kappa| r$. Since this spherical wave is the envelope of the full MTO
we see that, for negative $\kappa^2$, the MTO has a tail that decays exponentially with  
distance from its host site. The decay rate is controlled directly by the magnitude of $\kappa$.
This feature makes it suitable for tight-binding calculations.

Inside the MT spheres, the envelope spherical wave is augmented (i.e. replaced) with a solution 
of the \Schrodinger equation for the MT potential. The augmentation process is similar to the
augmentation of the plane wave in the Augmented Plane Wave (APW) method.\cite{loucks1967,singh1994}
The difference, of course, is that here we are augmenting a spherical wave instead of a plane wave. 
Note that the basis functions in the APW method are not suitable for tight binding since their
plane-wave envelopes are not localized in space. 

As we approach the host MT sphere from the outside, the basis function is determined
by Eq.(\ref{interstitial}). Note that, relative to the origin at the host site, this function 
is separable, i.e. it is a product of a radial and an angular part. 
Inside the host MT sphere, this function is augmented with a solution of the \Schrodinger
equation for the MT potential. Since this potential is spherically symmetric inside the MT, 
the basis function inside the host MT is also separable:

\begin{equation}
\phi_{i \alpha}(i \alpha | {\bf r}) = \big[ A \varphi_{i \alpha l} 
(e_{nl},r_{i \alpha}) 
+ B \dot{\varphi}_{i \alpha l}(e_{nl},r_{i \alpha}) \big] 
\cdot i^l Y_{lm}(\hat{\bf r}_{i \alpha})
\cdot \Theta(i \alpha | \boldr ),
\label{host_sphere}
\end{equation}

\noindent where the MT masking function $\Theta(i \alpha | \boldr )$ is equal to 1 if $\boldr$ is
inside the MT sphere $(i \alpha)$ and 0 otherwise. The energy dependence of the radial solution 
has been approximated with a linear combination of the radial solution $\varphi_{i \alpha l}$
and its energy derivative $\dot{\varphi}_{i \alpha l}$ 
($\equiv \partial \varphi_{i \alpha l} / \partial
\varepsilon$) calculated at a fixed chosen energy $e_{nl}$.\cite{andersen1975}
The coefficients are obtained by requiring the function to match continuously and smoothly with
the interstitial function at the MT radius:

\begin{equation}
\left[
\matrix{ \varphi_{i \alpha l} (s_{i \alpha}) & \dot{\varphi}_{i \alpha l} (s_{i \alpha}) \cr 
\varphi_{i \alpha l}' (s_{i \alpha}) & \dot{\varphi}_{i \alpha l}' (s_{i \alpha})}
\right] \cdot
\left[ \matrix{ A \cr B} \right] = 
\left[ 
\matrix{ K_l(\kappa, s_{i \alpha}) \cr K_l'(\kappa, s_{i \alpha})}
\right].
\end{equation}

\noindent Here a prime denotes differentiation with respect to the radial coordinate:
$\varphi' \equiv \partial \varphi / \partial r$.

The augmentation process at the remote spheres proceeds rather differently from the one
for the host sphere. As we approach a remote MT sphere from the interstitial region, the 
basis function is again determined by Eq.(\ref{interstitial}). The important difference 
from the host-sphere case is that relative to the origin at the remote site the interstitial
function is {\it not} separable. We can, however, express the spherical Hankel function 
at the interstitial region as a sum of separable functions in the neighbourhood of the remote 
site:\cite{danos1965,nozawa1966,talman1968,andersen1971a,gonis2000}

\begin{equation}
K_l(\kappa,r_{i \alpha}) \cdot i^l Y_L(\hat{\bf r}_{i \alpha})
= \sum_{L'} J_{l'}(\kappa,r_{i' \alpha'}) \cdot i^{l'} Y_{L'}(\hat{\bf r}_{i' \alpha'})
\cdot S_{L'L}(\kappa,\boldR_{i' \alpha'} - \boldR_{i \alpha}), \qquad
r_{i'\alpha'} < | \boldR_{i'\alpha'} - \boldR_{i \alpha} |,
\label{addition_theorem}
\end{equation}

\noindent with the spherical Bessel function as the radial part in each term:

\begin{equation}
J_l(\kappa,r) = \kappa^{-l} j_l(\kappa r)
= {1 \over z} \times \cases{\sinh z, &$l=0$, \cr |\kappa|^{-1} 
(\cosh z - z^{-1} \sinh z), &$l=1$, \cr
|\kappa|^{-2} [(3z^{-2} + 1)\sinh z - 3 z^{-1} \cosh z], &$l=2$.}
\end{equation}

\noindent The addition theorem for the Bessel functions, Eq.(\ref{addition_theorem}), is
a consequence of the fact that both sides of the identity are solution of the 
translationally-invariant Helmholtz equation. The spherical Hankel function, $K_l(\kappa,r)$, 
is regular everywhere except at the origin, while the spherical Bessel function, $J_l(\kappa,r)$,
is regular everywhere except at infinity. The domain of validity of Eq.(\ref{addition_theorem})
is just the domain where both sides of the identity are regular. 

The radial part of each term in the one-center expansion, Eq.(\ref{addition_theorem}),
is centered at the remote site $\boldR_{i' \alpha'}$. Thus the $(\varphi,\dot{\varphi})$ 
augmentation of the interstitial function, Eq.(\ref{interstitial}), at this remote site 
can be performed by augmenting each $J_l(\kappa,r_{i'\alpha'})$ with the radial part of 
the solution of the \Schrodinger equation for the spherically-symmetric MT potential centered
at $\boldR_{i'\alpha'}$. Thus we replace:

\begin{equation}
J_l(\kappa,r_{i'\alpha'}) \rightarrow \big[ C_l \varphi_{i'\alpha'l}(e_{nl},r_{i'\alpha'})
+ D_l \dot{\varphi}_{i'\alpha'l}(e_{nl},r_{i'\alpha'}) \big].
\end{equation}

\noindent The coefficients are again obtained by requiring continuity in the value of the
function and its first derivative with those of the interstitial envelope function at the
MT radius:

\begin{equation}
\left[
\matrix{ \varphi_{i' \alpha' l} (s_{i' \alpha'}) & \dot{\varphi}_{i' \alpha' l} (s_{i' \alpha'}) \cr 
\varphi_{i' \alpha' l}' (s_{i' \alpha'}) & \dot{\varphi}_{i' \alpha' l}' (s_{i' \alpha'})}
\right] \cdot
\left[ \matrix{ C_l \cr D_l} \right] = 
\left[ 
\matrix{ J_l(\kappa, s_{i' \alpha'}) \cr J_l'(\kappa, s_{i' \alpha'})}
\right].
\end{equation}

The part of basis function inside a remote MT sphere is thus given by:

\begin{equation}
\phi_{i \alpha}(i' \alpha' | {\bf r}) = \Theta(i' \alpha' | \boldr ) \cdot 
\sum_{L'}^{l \leq l_m} 
\big[ C_{l'} \varphi_{i'\alpha'l'}(r_{i'\alpha'})
+ D_{l'} \dot{\varphi}_{i'\alpha'l'}(r_{i'\alpha'}) \big]
\cdot i^{l'} Y_{L'}(\hat{\bf r}_{i' \alpha'})
\cdot S_{L'L}(\kappa,\boldR_{i' \alpha'} - \boldR_{i \alpha}).
\label{remote_sphere}
\end{equation}

\noindent The summation on the r.h.s. of the equation ideally runs to infinity. Practically,
a converged total energy is achieved using $l \leq l_m \sim 6-8$ in most cases.\cite{wills2000}
Note that, inside the remote sphere, an angular-momentum expansion is necessary since the
resulting function needs to match the envelope interstitial function at its MT sphere. Relative
to this remote site, the envelope function contains multiple harmonics since
it is centered at a different site.

The MT orbital, constructed from Eqs.(\ref{interstitial}), (\ref{host_sphere}), 
and (\ref{remote_sphere}), is continuous and smooth everywhere. Note, however, that it is 
not normalized and it does {\it not} have a separable form:

\begin{equation}
\phi_{i \alpha}({\bf r}) \neq u(r_{i \alpha}) Y_{lm}(\hat{\bf r}_{i \alpha}),
\end{equation}

\noindent in other words, the MTO does not have good angular-momentum quantum numbers.
Although the parts of the MTO inside the host MT and at the interstitials
are separable, the presence of the remote sites breaks the perfect spherical symmetry 
around the host site, and therefore also removes the separability. The MTO, however, does
obey exactly the point symmetry of its host site. Moreover, since the non-separable
contributions come from the remote sites, where the amplitude of the exponentially-decaying
envelope function is small, to a good approximation the MTO does transform as if 
it has the quantum numbers $(lm)$ of the part of the orbital inside the host sphere. 
The fact that the MTO is not fully separable should be
noted especially if one intends to map the FPLMTO result into a TB form using the Slater-Koster 
parametrization method. In the SK-LCAO parametrization method the basis functions
are assumed to be atomic orbitals; these orbitals are separable since they are derived from a 
single-atom spherically-symmetric potential.\cite{slater1954,slater1965,sharma1979,harrison1989} 

%% Section: Fourier Transform

\subsection{Fourier Transform}
\label{fourier_transform}

In applications of the FPLMTO method to ideal crystalline systems, instead of using the basis 
functions in real space, one works with the Bloch functions:

\begin{equation}
\psi_\alpha (\boldk, \boldr) = {1 \over \sqrt{N}} 
\sum_i \exp(i \boldk \cdot \boldR_i ) \ \phi_{i \alpha}(\boldr).
\label{bloch_function}
\end{equation}

\noindent The energy bands are obtained by solving the eigenvalue equation:

\begin{equation}
\det | H_{\alpha \beta} (\boldk) - E S_{\alpha \beta}(\boldk) | = 0.
\label{nonorthogonal_eigenvalue}
\end{equation}

\noindent The rank of the matrices is equal to the total number of the orbitals used for
all the atoms within the unit cell. The hamiltonian matrix is given by integration over
an arbitrarily chosen unit cell $\boldR_n$:

\begin{equation}
H_{\alpha \beta} (\boldk) = \langle \psi_\alpha (\boldk, \boldr) | 
H | \psi_\beta(\boldk,\boldr) \rangle
= {1 \over N} \sum_{ij} e^{i \boldk \cdot (\boldR_j - \boldR_i)} 
\sum_n \Big[ \langle \phi_{i \alpha} ( I_n) | \ H \ | 
\phi_{j \beta} ( I_n ) \rangle +
\sum_{\gamma} \langle \phi_{i \alpha} ( n \gamma ) | \ H \ | 
\phi_{j \beta} (n \gamma ) \rangle \Big],
\label{hamiltonian_matrix}
\end{equation}

\noindent and similar expression is obtained for the overlap matrix by replacing the hamiltonian
operator $H$ with identity. The vectors have been defined by:

\begin{equation}
\langle \boldr | \phi_{i \alpha} (n \gamma) \rangle = \phi_{i \alpha} (n \gamma | \boldr),
\end{equation}

\noindent which is the part of the basis function, centered at the host site $i \alpha$,
inside the MT sphere $n \gamma$ or the interstitial region $I_n$, which is defined to be 
the interstitial region within the unit cell $\boldR_n$.
Cross terms are absent from the square bracket in Eq.(\ref{hamiltonian_matrix}) 
since the overlap of their masking functions is zero. 

It should be mentioned that in the actual FPLMTO formalism, the matrix elements are not
computed using a multicenter expansion (over all lattice sites) as shown in Eq.(\ref{bloch_function}). 
Rather, using the addition theorem for the tail of the MT orbital, Eq.(\ref{addition_theorem}),
the multicenter expansion is first transformed into a one-center expansion (over angular momenta
at a single site), and this sum is the one that is actually computed.
\cite{skriver1984,wills2000} This, however, is merely a computational contrivance, albeit a pivotal 
one in the FPLMTO method, and should not mask the true structure of the basis function in 
{\bf k}-space as a simple Bloch sum of the MT orbitals.

In the FPLMTO method of Wills {\it et al.}\cite{wills2000}, the fact that each term of 
the matrix elements is to be calculated only over a certain region, either inside the MT sphere or
the interstitial region as seen in Eq.(\ref{hamiltonian_matrix}), is used to facilitate 
its computation. Instead of using the true MTO, which is complicated, one uses a pseudo 
basis orbital.\cite{weinert1980} This orbital is equal to the true MTO inside the relevant 
integration region but outside the region it is replaced with a smooth function which is chosen
to facilitate the computation. The integration can then be performed
efficiently by working in the momentum space.   

Using the translational properties of the Bloch functions, the matrix element can be written as:

\begin{equation}
H_{\alpha \beta} (\boldk) = 
\sum_j \exp(i \boldk \cdot \boldR_j ) \ H_{\alpha \beta} (\boldR_j),
\label{k_from_r}
\end{equation}

\noindent with the matrix elements in real space:

\begin{equation}
H_{\alpha \beta}(\boldR_j) = 
\sum_n \Big[ \langle \phi_{0 \alpha} ( I_n) | \ H \ | 
\phi_{j \beta} ( I_n ) \rangle +
\sum_{\gamma} \langle \phi_{0 \alpha} ( n \gamma ) | \ H \ | 
\phi_{j \beta} (n \gamma ) \rangle \Big].
\end{equation}

To calculate the inverse transform of Eq.(\ref{k_from_r}) we use the fact that the 
real-space matrix element $H_{\alpha \beta}(\boldR)$ decays exponentially with 
the distance $|\boldR|$. It is thus reasonable to ignore matrix elements between orbitals separated
by a distance greater than a certain cutoff radius $R_c$. In practice this can be performed 
by working with a finite crystal which contains $N_x$ unit cells along the $x$-direction such that
$N_x a_x > 2 R_c$, with $a_x$ being the lattice constant along the $x$-direction, and 
similar ranges for the $y$ and $z$ directions. 
The real-space matrix elements can then be obtained from a discrete Fourier transform:

\begin{equation}
H_{\alpha \beta} (\boldR_j) = 
{1 \over N} \sum_\boldk \exp(-i \boldk \cdot \boldR_j ) \ H_{\alpha \beta} (\boldk).
\label{r_from_k}
\end{equation}

\noindent Here $N = N_x N_y N_z$ and $k_i = \pi n_i/(N_i a_i)$ with 
$-N_i < n_i \leq N_i$ for $i \in \{ x, y, z \}$. 

Some practical remarks should be mentioned regarding this calculation of the  
inverse Fourier transform. First, the output matrix elements from the FPLMTO are calculated 
using the spherical harmonic $i^l Y_{lm}(\unitr)$ as the angular part of the basis function 
inside its host sphere. Usually it is easier to work with real basis functions and this
can be achieved by making linear combinations of the spherical harmonics.\cite{weissbluth1978}
Second, the basis functions are not normalized. The normalized matrix elements are obtained 
by using the overlap matrix elements:

\begin{equation}
H_{\alpha \beta}'(\boldR) = S_{\alpha \alpha} (\bold0)^{-{1 \over 2}} \cdot
H_{\alpha \beta} (\boldR) \cdot S_{\beta \beta} (\bold0)^{-{1 \over 2}}.
\end{equation}

\noindent Third, the actual displacement vector connecting the centers of the two orbitals in 
$H_{\alpha \beta} (\boldR)$ is not $\boldR$ but is instead $(\boldR + \boldtau_\beta -
\boldtau_\alpha)$. Instead of the original hamiltonian matrix in Eq.(\ref{k_from_r}), it is 
preferable to work with a modified matrix that is related to the original by a unitary
transformation:

\begin{equation}
H_{\alpha \beta}''(\boldk) = e^{i \boldk \cdot (\boldtau_\beta - \boldtau_\alpha)}
H_{\alpha \beta}(\boldk) = \sum_j e^{i \boldk \cdot (\boldR_j + \boldtau_\beta - 
\boldtau_\alpha)} \ H_{\alpha \beta} (\boldR_j).
\label{hopping}
\end{equation}

\noindent This positions the origin at the center of orbital $(0\alpha)$ and allows the 
matrix elements to obey the point symmetry of the corresponding site. It may therefore 
be used to reduce the number of matrix elements that need to be stored. Note that the 
eigenvalues are unaffected by this unitary transformation.

\begin{figure}
      \epsfysize=50mm
      \centerline{\epsffile{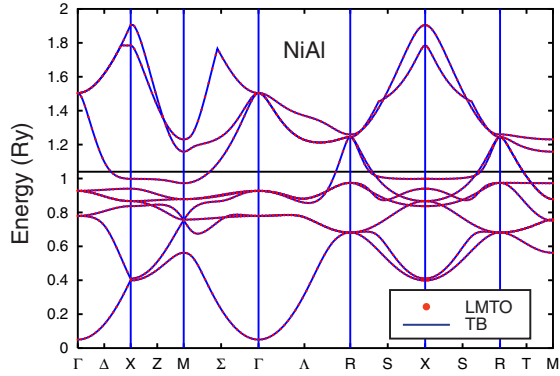}}
\bigskip
\caption{Energy bands for NiAl (only the 9 lowest bands are shown). The lines are the output 
eigenvalues from FPLMTO while the dots are calculated using the TB parameters. The two sets 
coincide with each other since the TB parameters are obtained by direct extraction from the FPLMTO.
The horizontal line is drawn at the Fermi energy, $E_F$ = 1.0417 Ry.} 
\label{NiAl_bands}
\end{figure}

%% Section: Lowdin Transformation

\subsection{L\"owdin Orthogonalization}
\label{lowdin_transform}

The real-space hamiltonian matrix elements in Eq.(\ref{r_from_k}) (together with the overlap 
matrix elements $S_{\alpha \beta}$ which can be obtained similarly) can be used directly in 
a nonorthogonal recursion calculation.\cite{riedinger1989,ballentine1986,mckinnon1995} 
Unfortunately, this requires us to obtain the inverse of the overlap matrix in real space, which is 
a non-trivial computational task. In this paper we instead use an orthogonal hamiltonian which is 
obtained by L\"owdin orthogonalization of the original hamiltonian and overlap matrices.
\cite{weissbluth1978} An alternative scheme is to use the chemical pseudopotential approach
and work with a non-hermitian matrix $S^{-1}H$.\cite{bullett1980,foulkes1993} The advantage
of L\"owdin orthogonalization is that it is a symmetry transformation: angular symmetry is preserved, 
i.e., orbitals with angular momentum $l$ are not mixed by the transformation with orbitals 
with $l' \neq l$. \cite{altmann1995} This means that the real-space matrix elements can still
be parametrized by Slater-Koster parametrization, e.g., using the procedure proposed by McMahan and 
Klepeis.\cite{mcmahan1997} 

Since we are working with a perfect crystal, the L\"owdin orthogonalization process can be applied 
to the momentum-space matrix elements. One first diagonalize the overlap matrix:

\begin{equation}
D=U^\dagger \cdot S \cdot U,
\end{equation}

\noindent where $U$ is the matrix containing the eigenvectors of $S$ as its columns, 
and $D$ is a diagonal matrix containing the eigenvalues of $S$. Using these matrices, we define
a new matrix:

\begin{equation}
A = U \cdot {D^{-{1 \over 2}}} \cdot U^\dagger.
\end{equation}

\noindent Since the overlap matrix is a positive-definite hermitian matrix, its eigenvalues
are all positive real numbers and the matrix $D^{-{1 \over 2}}$ is well defined. In practice,
however, small negative eigenvalues of $S$ may sometimes appear due to the finite machine 
precision used in the computation. This problem is especially relevant if the diagonalization
of the overlap matrix is performed in real space and when one uses an overlap matrix which is 
necessarily {\it approximate} due to the finite cutoff imposed on the range of the matrix elements
or due to the Slater-Koster two-center approximation used in parametrizing them. One 
commonly-used fix to this problem is to add a diagonal matrix with small elements to make 
the overlap matrix positive definite.\cite{roder1997} The orthogonal L\"owdin hamiltonian matrix
is obtained by sandwiching the original hamiltonian matrix with $A$ matrices:

\begin{equation}
H^{(L)} = A \cdot H \cdot A.
\end{equation}

\noindent The nonorthogonal eigenvalue problem, Eq.(\ref{nonorthogonal_eigenvalue}), is 
then transformed to an equivalent orthogonal one:

\begin{equation}
\det | H_{\alpha \beta}^{(L)} (\boldk) - E \delta_{\alpha \beta}(\boldk) | = 0.
\label{orthogonal_eigenvalue}
\end{equation}

\noindent The orthogonal hopping parameters are obtained by substituting 
$H_{\alpha \beta}^{(L)} (\boldk)$ in Eq.(\ref{r_from_k}). These are the parameters that 
we use in the recursion calculation which is described in the rest of this paper.

\section{Recursion Calculation}
\label{recursion}

The recursion method, initially introduced to electronic-structure calculation by Haydock,
is one of the most efficient and stable ways of calculating the electronic Green's function
from the TB hamiltonian describing the system:\cite{haydock1980,pettifor1985} 

\begin{equation}
G_{uv}(E) = \langle u | (E- H)^{-1} | v \rangle,
\end{equation}

\noindent where $|u \rangle$ and $|v \rangle$ are two arbitrary vectors in the Hilbert space
for the problem. The local density of states (LDOS) is a projection of the total density of states 
to a local orbital:

\begin{equation}
\rho_\alpha(E) = \sum_n | \langle \alpha | \Psi_n \rangle |^2 \ \delta(E - E_n),
\end{equation}

\noindent where $| \Psi_n \rangle$ is an eigenvector with eigenvalue $E_n$. This can
be obtained from the Green's function:

\begin{equation}
\rho_\alpha(E) = - {1 \over \pi} \ {\rm Im} \ \big[ G_{\alpha \alpha}(E + i \epsilon) \big],
\end{equation}

\noindent with $\epsilon \rightarrow 0^+$. The token Im$[\cdots]$ denotes the operation of taking 
the imaginary part of its argument. The recursion method outputs the Green's function
as a continued fraction:

\begin{equation}
G(E) = {1 \over (E - a_0) - {\textstyle b_1^2 \over {\textstyle (E - a_1) - 
{\textstyle b_2^2 \over {\textstyle \hphantom{E} \cdots \hphantom{E} }    }}}}. 
\label{greens_function}
\end{equation}

\noindent Note that there are two conventions used in the literature regarding the indexing 
of the $a$-parameters in the above continued fraction. One convention is to assign $a_0$ as the
first $a$-parameter,\cite{haydock1980,ballentine1986} while the other use $a_1$ as the first
parameter.\cite{turchi1982} Here we follow the first convention.
The calculation of the LDOS is therefore reduced to the computation of the strings of
recursion coefficients $\{ a_n \}$ and $\{ b_n \}$. The procedure for performing this has been
described extensively in the literature and will not be repeated here.\cite{haydock1980,pettifor1985}

\begin{figure}
      \epsfysize=50mm
      \centerline{\epsffile{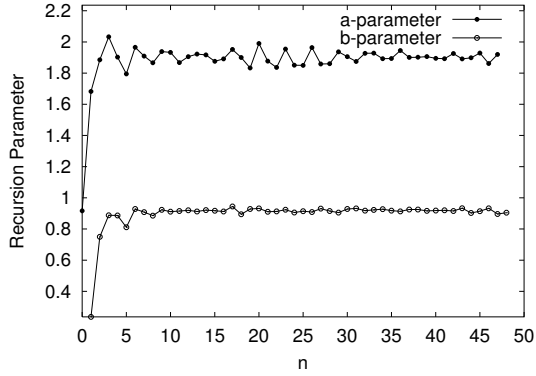}}
\bigskip
\caption{Recursion parameters for the LDOS of Ni-$d$ orbitals. Filled circles ($\bullet$) 
are the $a_n$ parameters while open circles ($\circ$) are $b_n$ parameters.} 
\label{Ni_d_parameters}
\end{figure}

\begin{figure}
      \epsfysize=50mm
      \centerline{\epsffile{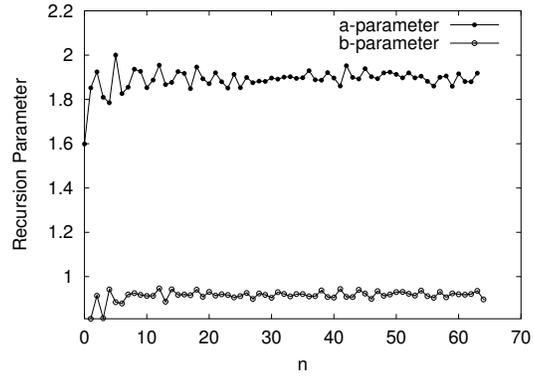}}
\bigskip
\caption{Recursion parameters for the LDOS of Al-$p$ orbitals. Filled circles ($\bullet$) 
are the $a_n$ parameters while open circles ($\circ$) are $b_n$ parameters.} 
\label{Al_p_parameters}
\end{figure}

We have used the recursion method to calculate the LDOS for NiAl crystal. This
alloy crystallizes in B2 structure with two atoms per (cubic) unit cell. One component (Ni or Al) sits
at $(0,0,0)$, while the other is at the body-center of the cube, $( {1 \over 2}, {1 \over 2},
{1 \over 2})$. We obtained the real-space orthogonal hopping parameters by direct extraction 
and L\"owdin transformation as described in the preceding section. The parameters were 
extracted from a single-$\kappa$ FPLMTO calculation using 9 $spd$ orbitals and 
$\kappa^2 = -0.04$ Ry for each Ni or Al atom. This rather small value of $\kappa$ produces basis 
functions with relatively long range (the envelope of the basis function decays roughly as 
$e^{-|\kappa| r}$). Using a larger negative value of $\kappa$, however, produces a greater 
discrepancy between the output bands from this single-$\kappa$ calculation and the corresponding
accurate bands from a multiple-$\kappa$ calculation. The chosen value of $\kappa$ that we used
is obtained by compromising the need to have a short-ranged TB basis functions with the 
requirement to have a precision comparable to accurate multiple-$\kappa$ result. 

\begin{figure}
      \epsfxsize=178mm
      \centerline{\epsffile{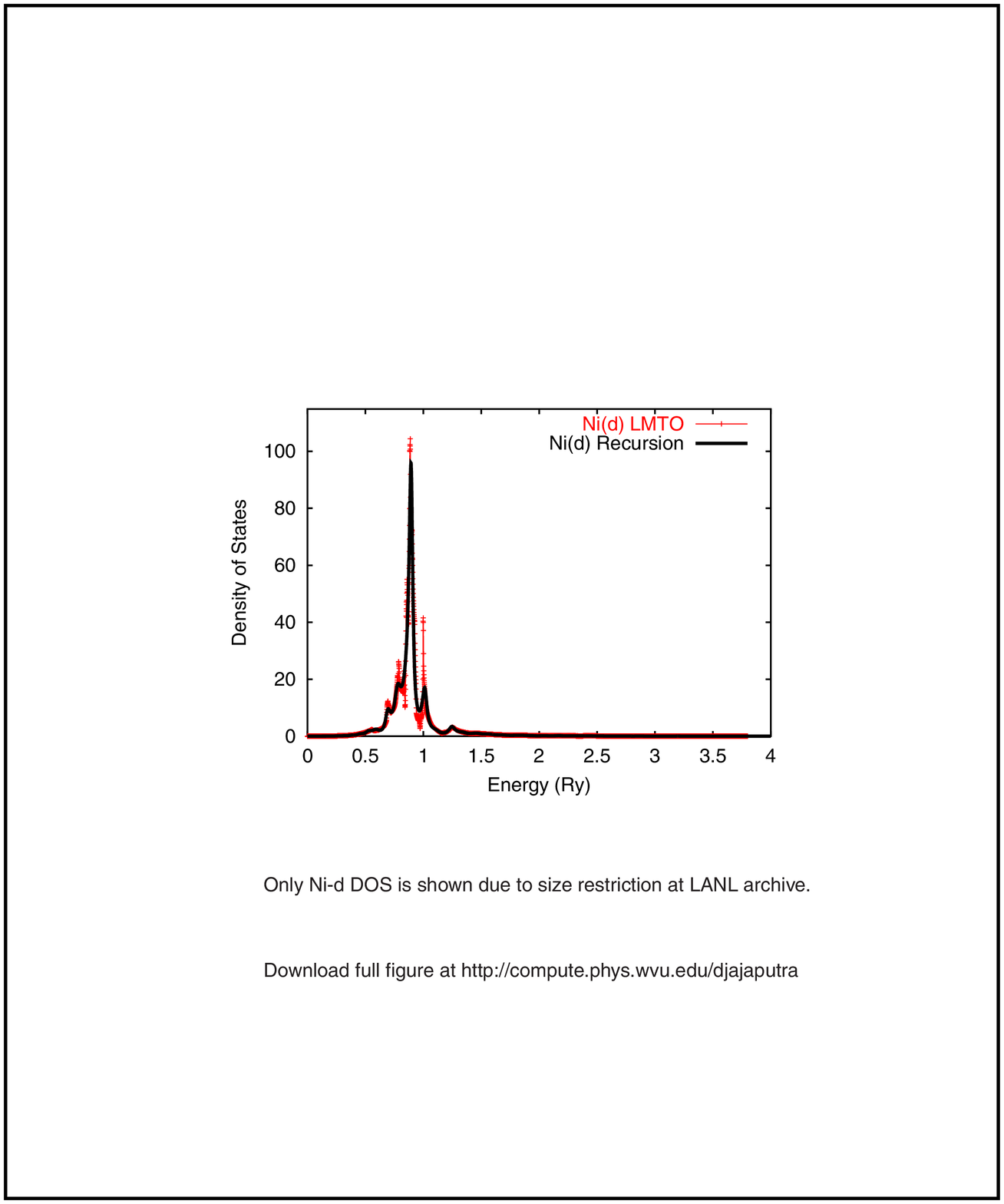}}
\bigskip
\caption{The local density of states (LDOS) for Ni and Al states in NiAl as calculated
using the tight-binding recursion method (dark full line) and from Brillouin-zone integration
of FPLMTO eigenvalues with atom and angular-momentum projection (grey line with plus (+) signs).
(a,b,c) The left column shows $s,p,d$ LDOS's for Ni; (d,e,f) the right column displays the 
$s,p,d$ LDOS's for Al. The Fermi energy is at $E_F$ = 1.0417 Ry.} 
\label{alldos}
\end{figure}

All of the extracted parameters are saved in a database for use in the recursion calculation 
(crystal symmetry was used to reduce the size of the database). They were calculated for separation 
distance of up to 8 lattice constants. The obtained TB parameters 
fall roughly one order of magnitude for each increase in separation distance of 
one lattice constant. The recursion calculation was performed using a cubic cluster of 
$(16)^3$ unit cells and the LDOS's were computed for Ni and Al orbitals 
within the unit cell at the center of the cluster. Since the TB parameters connecting the 
orbitals at the central unit cell to the ones at the surface of the cluster are practically zero,
the calculated LDOS should not depend on the size of the cluster if we
increase it further.
 
Figs.~\ref{Ni_d_parameters} and \ref{Al_p_parameters} show the calculated recursion parameters
for Ni-$d$ and Al-$p$ states, respectively. Here the $d$-state is defined by:

\begin{equation}
| d \rangle = {1 \over \sqrt{5}} \sum_{m=-2}^2 |l=2,m \rangle,
\end{equation}

\noindent with a corresponding definition for the $p$-state. It is seen that the parameters
tend to settle around certain constant values. Indeed it is known that as $n \rightarrow \infty$:
\cite{turchi1982}

\begin{equation}
a_n \rightarrow (E_t + E_b)/2, \quad b_n \rightarrow W/4,
\end{equation}

\noindent where $E_t$ and $E_b$ are the values of energy at the top and bottom of the spectrum,
respectively, and $W = (E_t - E_b)$ is the total bandwidth. Furthermore, asymptotically the 
parameters should oscillate around their 
limit values with a decaying amplitude as $n \rightarrow \infty$.\cite{gaspard1973} 
The frequency and the rate of decay are determined by specific features (the Van Hove 
singularities) of the spectrum.\cite{hodges1977}   

The recursion parameters shown in Figs.~\ref{Ni_d_parameters} and \ref{Al_p_parameters} are seen
to roughly follow this decayed oscillation prediction. However, for large values of $n$ (greater
than $n_m \sim 40$), the amplitudes of the deviation tend to increase rather than decrease. 
We interpret this as the limit at which the machine precision noise start to interfere with 
the calculation. This interpretation is supported by our experience that incorporating the 
recursion parameters with $n > n_m$ in the reconstruction of LDOS in general does not help improve
the agreement, and indeed it tends to reduce the agreement, with the accurate LDOS obtained 
from the FPLMTO method. Note that the main workhorse in the recursion method involves a matrix-vector
multiplication with the dimension of the vector being the total number of orbitals in the cluster.
In our case this dimension is equal to 73728 (= $16^3$ unit cells $\times$ 2 atoms/cell $\times$
9 orbitals/atom). Compounded with this large dimension is the fact that the elements of the matrix 
and the vector in general have widely different orders of magnitude which makes the calculation 
rather prone to rounding errors.

Figs.~\ref{alldos} display the main results in this paper: the {\it s,p,d} LDOS
of Ni and Al in NiAl as calculated using the TB recursion method. These LDOS's are compared with 
the corresponding accurate spectra obtained from Brillouin zone integration using atom and 
angular-momentum projection of the eigenvalues obtained directly from the FPLMTO method. 
The agreement in general is good although the TB recursion method is unable to precisely 
reproduce the sharp peaks in the FPLMTO spectra. Not very surprisingly, the best fit is 
obtained for the tightly-bound Ni-$d$ state.

The limited number ($n_m \sim 40$) of reliable recursion parameters that we can use here is generally
not sufficient for acceptable reproduction of the FPLMTO LDOS. Furthermore, construction 
of LDOS by directly taking the imaginary part of the Green's function, Eq.(\ref{greens_function}),
normally introduces an undesirable feature in the output LDOS in the form of spurious 
rapid oscillations near the edges of the spectrum, as can still be seen in, e.g., 
Fig.~\ref{alldos}a. The simplest extrapolation method for the recursion parameters is
the square-root terminator method:\cite{haydock1980} one simply replaces all $a_n$ and $b_n$ 
for $n > n_m$ with constants $a_\infty$, and $b_\infty$. The tail of the continued fraction for 
the Green's function can then be summed exactly as a square-root expression. The best values
for the terminating coefficients $a_\infty$ and $b_\infty$ can be obtained using the 
Beer-Pettifor method.\cite{beer1984}. Other ways to construct a terminator for the 
continued fraction have been suggested,\cite{haydock1985} but for our calculation we have
chosen to use the Beer-Pettifor method. Finally, further smoothing of the spectra shown 
in Fig.~\ref{alldos} have been performed using a Chebyshev-polynomial method.\cite{vargas1994}

\section{Summary}

In this paper, we have described a method for directly extracting real-space TB parameters from
a FPLMTO method. The basis functions used in the FPLMTO calculation and their construction are
described in considerable details. Special emphasis has been placed on the fact that these 
basis functions do not have a well-defined angular-momentum quantum number, although to a 
good approximation they do. We believe this fact should be kept in mind in transcribing
FPLMTO matrix elements to the Slater-Koster two-center form, since there it is implicitly
assumed that the basis functions do have a good angular-momentum quantum number.

This direct extraction method has been applied to an intermetallic alloy NiAl. The TB parameters
extracted have been used as input to a real-space calculation of local densities of states 
using the recursion method. We believe this is the first application of the direct extraction
method to a real-space calculation of the electronic structure of an intermetallic alloy. The
good agreement between the LDOS obtained using the TB extracted from the FPLMTO method and
the accurate LDOS obtained directly from the FPLMTO method shows the feasibility of using
this method to study more complicated cases, e.g., alloys with small concentration of 
impurity atoms.

This work was supported by AF-OSR Grant No. F49620-99-1-0274. DD would like to thank Dr.~M.J. Mehl
of the Naval Research Laboratory, Washington D.C., for useful discussions and insightful inputs 
regarding the NRL Tight-Binding code.

\end{document}